\documentclass[preprint]{ptephy_v1}

\preprintnumber{KEK-TH-2361, J-PARC-TH-0254} 
\usepackage{hyperref}
\usepackage{setspace,bm}

\newcommand{\up}{\uparrow}
\newcommand{\down}{\downarrow}
\newcommand{\be}{\begin{equation}}      
\newcommand{\ee}{\end{equation}}      
\newcommand{\bea}{\begin{eqnarray}}      
\newcommand{\eea}{\end{eqnarray}}

\newcommand{\cU}{{\cal U}}

\begin{document}

\title{Dissipation-induced dynamical phase transition in postselected quantum trajectories}


\author{Tomoya Hayata}
\affil{Departments of Physics, Keio University, 4-1-1 Hiyoshi, Kanagawa 223-8521, Japan \email{hayata@keio.jp}}

\author[2,3,4]{Yoshimasa Hidaka}
\affil[2]{KEK Theory Center, Tsukuba 305-0801, Japan}
\affil[3]{Graduate University for Advanced Studies (Sokendai), Tsukuba 305-0801, Japan}
\affil[4]{RIKEN iTHEMS, RIKEN, Wako 351-0198, Japan}

\author[5]{Arata Yamamoto}
\affil[5]{Department of Physics, The University of Tokyo, Tokyo 113-0033, Japan}


\begin{abstract}%
It is known that effects of dissipation or measurement backreaction in postselected quantum trajectories are described by a non-Hermitian Hamiltonian, but their consequences in real-time dynamics of many-body systems are yet to be elucidated.
Through a study of a non-Hermitian Hubbard model, we reveal a novel dissipation-induced dynamical phase transition in postselected quantum trajectories, where time controls the strength of postselection and becomes the intrinsic parameter inducing the phase transition.
Our findings are testable in ultracold atom experiments and may open a new avenue in the dissipative engineering of quantum systems.
\end{abstract}

\subjectindex{I22}

\maketitle

\section{Introduction} 
Understanding the dynamics of dissipative quantum many-body systems has been one of the key topics in recent years. 
Among them, two nonunitary dynamics are found to serve a new platform of physics: One is the conditional dynamics, which is obtained by post-selecting quantum trajectories from the unconditional Lindbladian dynamics of open quantum systems. The other is the monitored dynamics, which is a hybrid quantum system composed of the unitary evolution and repeated projective measurements. Intriguing phenomena such as the entanglement phase transition have been found so far~\cite{PhysRevB.99.224307,PhysRevX.9.031009,PhysRevB.100.134306,PhysRevB.101.104302}, but many-body phenomena unique to them are yet to be elucidated.

As aforementioned, theoretically, and also experimentally, the stochastically unraveling dynamics is  investigated instead of directly handling the quantum master equation. In this procedure, one solves the time evolution of a wave function  under some stochasticity instead of that of a density matrix.
If we employ the quantum trajectory method~\cite{Daley:2014fha,Breuer}, the dynamics is decomposed into two parts: One is the nonunitary evolution described by the Schr{\"o}dinger equation with an effective non-Hermitian Hamiltonian. The other is the quantum jump process, which is a stochastic loss event.
Although we can reconstruct the density matrix under the Lindbladian dynamics by averaging the loss event, or equivalently, the many trial wave functions, we restrict ourselves to follow the single trial wave function that experiences no loss event.
The dynamics of the constrained wave function leads us to study the non-Hermitian quantum mechanics (see, e.g., Ref.~\cite{Ashida:2020dkc} for a review).
However, genuine many-body physics is yet to be elucidated~\cite{PhysRevX.4.041001,2017NatCo...815791A,Lourenco:2018kvh,PhysRevLett.121.203001,PhysRevLett.123.090603,PhysRevB.102.081115,PhysRevB.102.235151,PhysRevLett.126.170503,PhysRevB.104.125102}, and in particular, it still lacks reliable methods for accurate large-scale numerical simulations.

\begin{figure}[t]
\begin{center}
 \includegraphics[width=.55\textwidth]{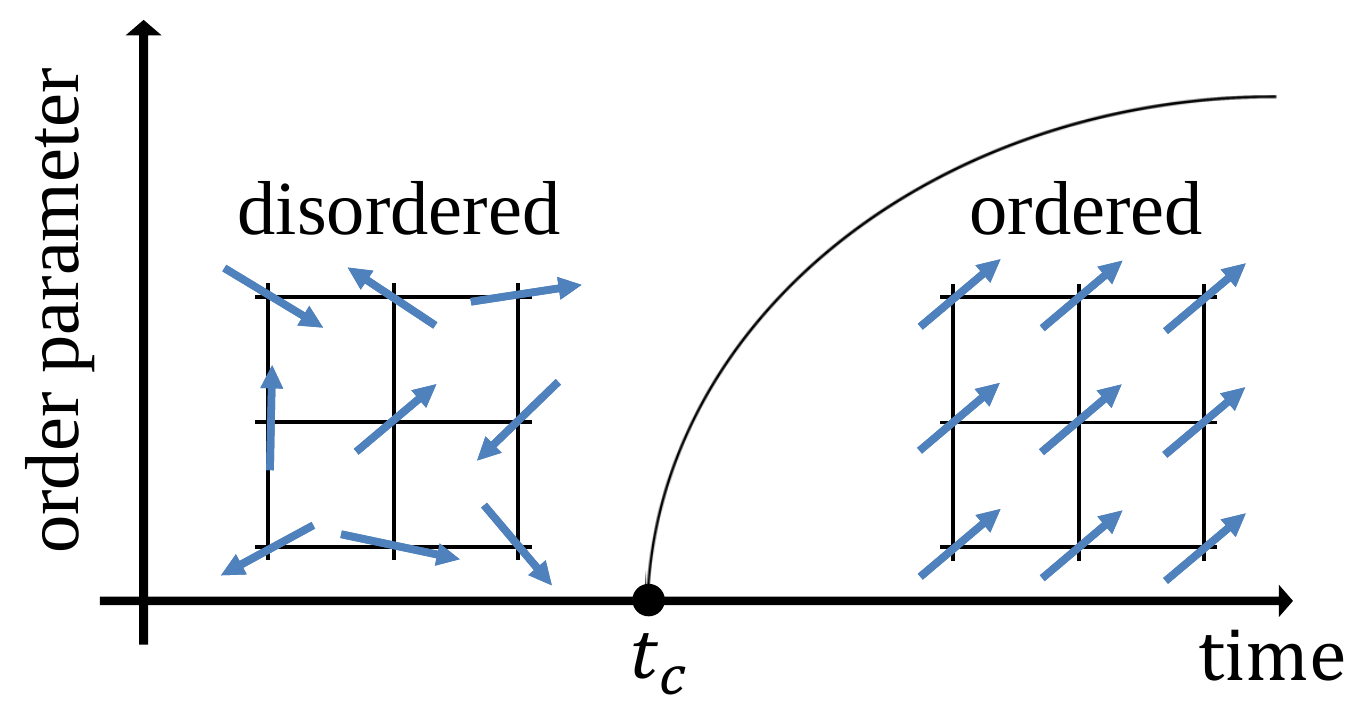}
\caption{
\label{fig1} Phase transition at finite time.
An order parameter of spontaneous symmetry breaking shows the characteristic behavior of the continuous phase transition as a function of time.
}
\end{center}
\end{figure}

In this paper, we reveal a dissipation-induced dynamical phase transition in postselected quantum trajectories. 
Using right- and left-eigenstates, and eigenvalues of a non-Hermitian Hamiltonian $|R_n\rangle$, $\langle L_n|$, and $E_n$, the constrained wave function at time $t$ reads
\bea
|\psi(t)\rangle
=\sum_{n}c_ne^{-it{\rm Re}[E_n]}e^{t{\rm Im}[E_n]}|R_n\rangle ,
\eea
where $c_n=\langle L_n|\psi(0)\rangle$.
Thus, eigenstates with a negatively larger imaginary part are exponentially eliminated by postselection as time evolves, and the long-time dynamics is governed by eigenstates with negatively smaller imaginary part, which results in the dramatic change of e.g., magnetism by dissipation {\it in the steady state}~\cite{PhysRevLett.124.147203}. 
Now a natural question arises: What happens {\it at finite time} with a large number of degrees of freedom? 
Since exponential suppression by postselection plays the same role as that by temperature in equilibrium systems, we expect a finite-time phase transition if an initial state is set away from the final steady state. Here, time controls the strength of postselection, and is  the intrinsic parameter inducing the phase transition,  as illustrated in Fig.~\ref{fig1}. 

To demonstrate that such a dynamic phase transition indeed occurs, we study the time evolution of a non-Hermitian Hubbard model in which the coupling strength of the Hubbard interaction is pure-imaginary. Such a non-Hermitian Hubbard model with complex coupling strength can be realized, e.g., in dissipative ultracold atoms, 
and its dynamical properties are intensively investigated~\cite{PhysRevLett.124.147203,2019QS&T,doi:10.1126/sciadv.1701513,PhysRevA.99.031601,PhysRevLett.123.123601,PhysRevLett.127.055301}.
We compute the time evolution of magnetic correlation functions after a quantum quench, and show that 
the system initially set in a symmetry-unbroken state suddenly turns into a symmetry-broken state during the time evolution. 
In contrast to the dynamical quantum phase transition in Hermitian systems~\cite{Heyl:2017blm}, an order parameter of spontaneous symmetry breaking exhibits non-analyticity at a critical time in the infinite volume limit (See Fig.~\ref{fig1}).

As a computational tool, we adopt a large-scale simulation of the fermionic quantum Monte Carlo.
Even though it is extremely challenging to compute the time evolution of a large quantum system exactly, the simulation is ab initio, unbiased, and applicable even to a higher-dimensional system.
This enables us to study many-body phenomena with a large number of degrees of freedom, and it plays an essential role in determining the existence of phase transitions, and the critical properties such as the universality class.

\section{Model and formulation} We consider two-component fermions on a three-dimensional cubic lattice $\bm r=(x,y,z)$. 
The unitary evolution is described by the free Hamiltonian,
\bea
H  = -w\sum_{{\bm r},j,\sigma}\left[ c^\dagger_{\bm r\sigma}c_{\bm r+\hat{j}\,\sigma}+c^\dagger_{\bm r+\hat{j}\,\sigma}c_{\bm r\sigma}\right] ,
\label{eq:free}
\eea
where $c^\dagger_{\bm r\up,\down}$ and $c_{\bm r\up,\down}$ are the creation and annihilation operators of the $\up$- and $\down$-component of fermions at a site $\bm r$, respectively. $w$ is a hopping parameter between the nearest neighbor sites, and $\hat{j}=\hat{x},\hat{y},\hat{z}$ is the unit lattice vector along the $j$ direction.
We consider the dissipative dynamics in the presence of particle loss due to inelastic collisions. When the $\up$- and $\down$-fermions occupy a site simultaneously, they acquire the kinetic energy from inelastic collisions and quickly escape from the system.
Such a loss process is described by the quantum master equation in the Lindblad form~\cite{Breuer},
\bea
\frac{d\rho}{dt} = -i\left(H\rho-\rho H\right)
+\sum_{\bm r}\gamma_{\bm r}\left( \Gamma_{\bm r} \rho \Gamma_{\bm r}^\dagger-\frac{1}{2}\Gamma_{\bm r}^\dagger\Gamma_{\bm r}\rho-\frac{1}{2}\rho\Gamma_{\bm r}^\dagger\Gamma_{\bm r} \right),
\label{eq:lindblad}
\eea
where $\rho$ is the density matrix of the system. The first and second terms determine the unitary and dissipative dynamics, respectively, where $\Gamma_{\bm r}$ is the quantum jump operator and $\gamma_{\bm r}$ is the strength of dissipation.
In our case, $\Gamma_{\bm r}$ removes the pair of fermions occupying the site $\bm r$ at rate $\gamma_{\bm r}$, so that the quantum jump operator is given as $\Gamma_{\bm r}=c_{\bm r\up}c_{\bm r\down}$. Since the loss rate is independent of the site, we take $\gamma_{\bm r}=2\gamma$, where the factor of $2$ is introduced for notational convenience. 

We employ the quantum trajectory method~\cite{Daley:2014fha}. Then, the dynamics is decomposed into the nonunitary evolution and quantum jump process.
By post-selecting the quantum trajectories, we follow the time evolution of the wave function that experiences no particle loss, which can be recovered from the non-Hermitian quantum mechanics described by 
\bea
i\frac{d}{dt}|\Psi(t)\rangle=H_{\rm eff}|\Psi(t)\rangle ,
\eea
and the non-Hermitian Hamiltonian,
\bea
H_{\rm eff} = -w\sum_{\bm r,j,\sigma}\left[ c^\dagger_{\bm r\sigma}c_{\bm r+\hat{j}\,\sigma}+c^\dagger_{\bm r+\hat{j}\,\sigma}c_{\bm r\sigma}\right]
-i\gamma\sum_{\bm r}  c^\dagger_{\bm r\up}c^\dagger_{\bm r\down}c_{\bm r\down}c_{\bm r\up}  .\;\;\;\;
\label{eq:nonhermite}
\eea
These define our model investigated in this paper.
As an initial condition, we consider the N{\'e}el state, that is, the half-filled state with even (odd) sites being occupied by the $\up$- ($\down$-) components.
Since $\langle\Psi(t)|\Psi(t)\rangle$ gives the persistent probability that no quantum jump process occurs, by taking the conditional probability into account, the expectation value of a physical observable $\hat{O}$ under the conditional dynamics is given as 
\bea
\langle \hat{O}\rangle(t) = \frac{\langle\Psi(t)|\hat{O}|\Psi(t)\rangle }{\langle\Psi(t)|\Psi(t)\rangle} .
\label{eq:expectation}
\eea
We compute it based on the sign-free auxiliary-field quantum Monte Carlo detailed in Appendix~\ref{sec:QMC}.

\section{Numerical simulation}
\begin{figure}[t]
\begin{center}
 \includegraphics[width=.55\textwidth]{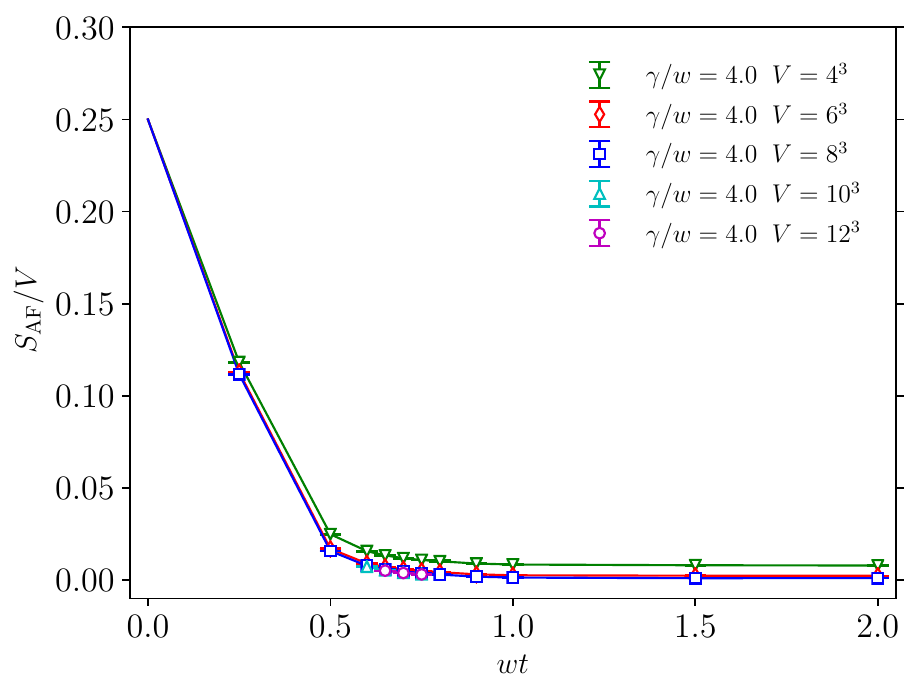}
\caption{\label{fig2}Time evolution of the anti-ferromagnetic spin structure factor. The initial state is the N{\'e}el state.
}
\end{center}
\begin{center}
 \includegraphics[width=.55\textwidth]{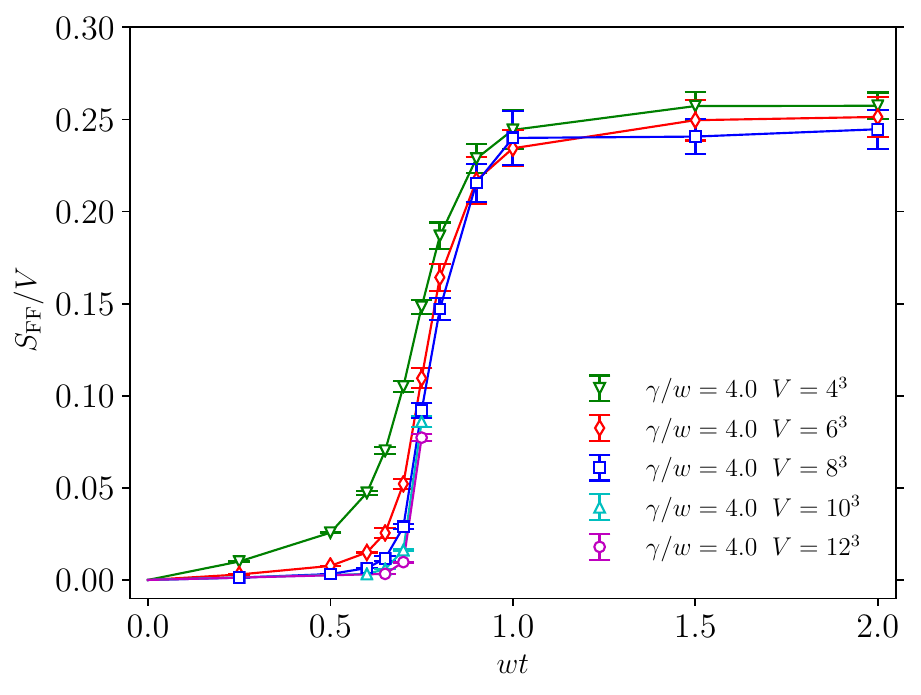}
\caption{\label{fig3}Time evolution of the ferromagnetic spin structure factor. The initial state is the N{\'e}el state.
}
\end{center}
\end{figure}

We compute the time evolution of the antiferromagnetic spin structure factor 
\bea
\frac{S_{\rm AF}}{V}=\left\langle \left[\frac{1}{V}\sum_{\bm r}(-)^{x+y+z}\bm S_{\bm r}\right]^2\right\rangle ,
\eea
and
the ferromagnetic spin structure factor 
\bea
\frac{S_{\rm FF}}{V}=\left\langle \left[\frac{1}{V}\sum_{\bm r}\bm S_{\bm r}\right]^2\right\rangle ,
\eea
where $\bm S_{\bm r}$ is the spin operator, given explicitly as
\bea
\bm S_{\bm r}=(S^x_{\bm r},S^y_{\bm r},S^z_{\bm r})=\frac12 (c^\dagger_{\bm r\up}c_{\bm r\down}+c^\dagger_{\bm r\down}c_{\bm r\up},-ic^\dagger_{\bm r\up}c_{\bm r\down}+ic^\dagger_{\bm r\down}c_{\bm r\up},c^\dagger_{\bm r\up}c_{\bm r\up}-c^\dagger_{\bm r\down}c_{\bm r\down}).
\eea
The presence of magnetic long-range orders can be judged by the nonvanishing of those structure factors in the large volume limit, that is, those are order parameters of spontaneous symmetry breaking.
We fixed the Trotter step with $\Delta t=0.05/w$.
We have checked the convergence of the numerical results by changing $\Delta t$.
We imposed periodic boundary conditions.
We show the results at $\gamma/w=4.0$ with $V=L^3=4^3,6^3,8^3,10^3$, and $12^3$ in Figs.~\ref{fig2} and~\ref{fig3}.
From Fig.~\ref{fig2}, we see that the antiferromagnetic correlation decays exponentially fast.
This indicates that the initial-state dependence is quickly lost after a very short time.
The insensitivity of $S_{\rm AF}/V$ against $V$ implies that the antiferromagnetic order is not due to spontaneous symmetry breaking, but just remnant of the initial N{\'e}el order.
A more interesting thing happens in the ferromagnetic spin structure factor in Fig.~\ref{fig3}.
We clearly see that the magnetic correlation suddenly changes from paramagnetic to ferromagnetic at a certain time.
The volume dependence suggests that this change can be regarded as a phase transition in the infinite volume limit; the system initially in a symmetry-unbroken state non-analytically falls into a symmetry-broken state at a phase transition time.
There is a sharp contrast with the conventional dynamical quantum phase transition~\cite{Heyl:2017blm}, which is probed by the non-analyticity of the Loschmidt echo, that is, the overlap between time-evolved and reference (usually initial) states. 
The dynamical phase transition here is accompanied by spontaneous symmetry breaking and probed by the magnetic correlation function.
Amazingly, it is just like the equilibrium phase transition, and can be probed by the experimental measurement used for the equilibrium magnetic phase transition.

\begin{figure}[t]
\begin{center}
 \includegraphics[width=.55\textwidth]{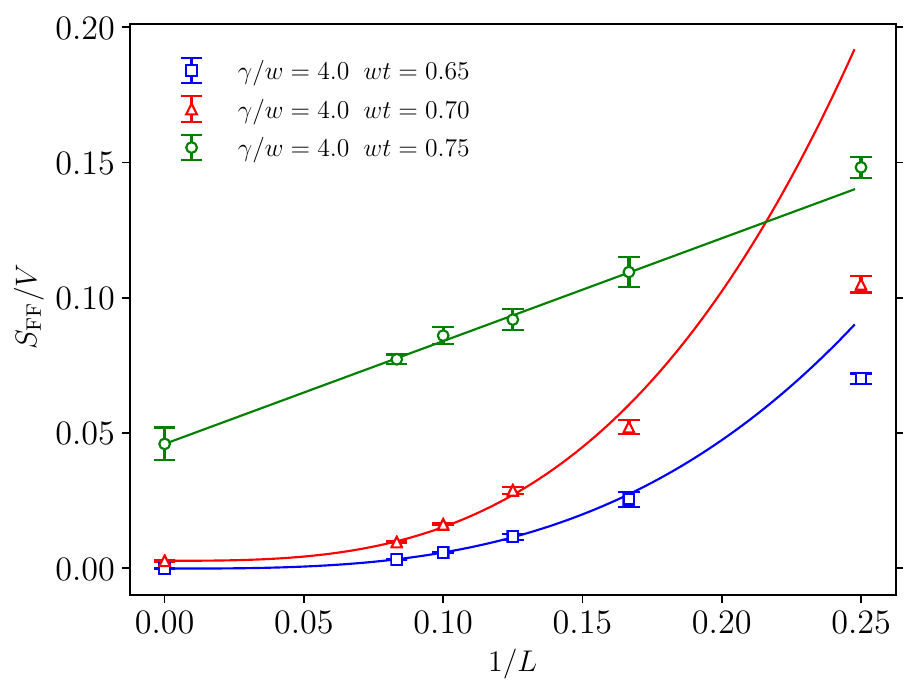}
\caption{\label{fig4}Finite-size scaling analysis near the critical time with $\gamma/w=4.0$. We fitted the data with $V=L^3=6^3,8^3,10^3$, and $12^3$ by $a+b/L$ ($wt=0.75$) or $a+b/L^3$ ($wt=0.65$ and $0.70$).
}
\end{center}
\begin{center}
 \includegraphics[width=.55\textwidth]{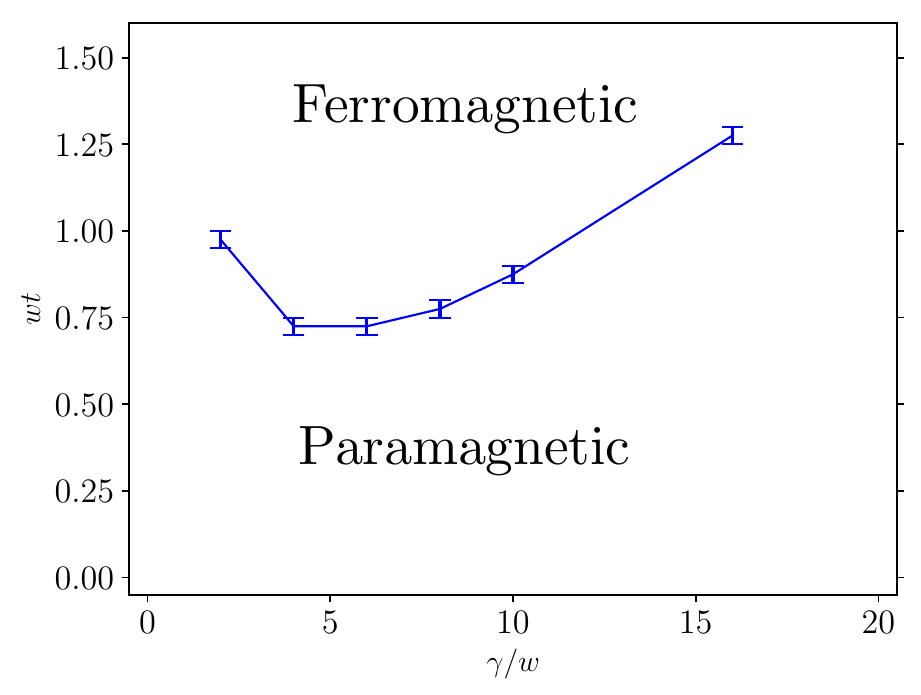}
\caption{\label{fig5}Magnetic phase diagram obtained from the finite-size scaling analysis of the quantum Monte Carlo data.
The curve is just for the eye guide.
}
\end{center}
\end{figure}

\begin{figure}[t]
\begin{center}
 \includegraphics[width=.55\textwidth]{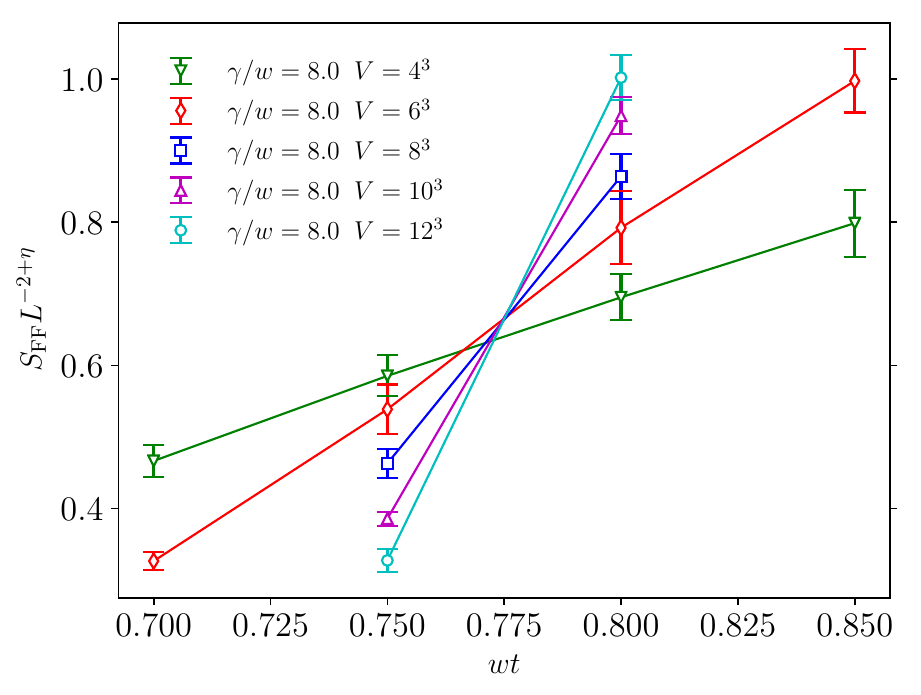}
\caption{\label{fig6} Finite-size scaling analysis at $\gamma/w=8.0$. The data crossing at one point indicates that scale invariance emerges in the phase transition point. 
}
\end{center}
\end{figure}

To quantify the dynamical phase transition, we performed the finite-size scaling analysis near the critical time as shown in Fig.~\ref{fig4} 
(The scale invariance emerges, and the phase transition time is indeed a critical point as detailed below).
After the critical time, the spin structure factor clearly shows the linear scaling and becomes nonzero in the infinite volume limit, while it shows the trivial volume-law scaling and goes to zero before the critical time. By fitting the data with $V=6^3,8^3,10^3$, and $12^3$, and extrapolating them to $V\rightarrow\infty$, we can estimate the critical time $t_c$; for $\gamma/w=4.0$, $t_c$ is in between $wt=0.70$ and $wt=0.75$ as seen in Fig.~\ref{fig4}.
Repeating the analysis for various $\gamma$, we draw the phase diagram of the dynamical phase transition in Fig.~\ref{fig5}.
As seen in Fig.~\ref{fig5}, $t_c$ becomes shorter at intermediate loss rate $\gamma/w\sim5$.
This is natural, because the hopping dynamics of free fermions dominates, and the system is expected to be paramagnetic at a weak $\gamma$, while the state remains as the initial state due to the quantum Zeno effect at a strong $\gamma$. 
We note that the absolute value of $t_c$ may depend on the initial state, as readily understood by choosing the state at $0<t<t_c$ as a new initial state.
We expect, however, that the existence of the phase transition and its critical property are universal as long as the initial state is a superposition of different total spin sectors including the pramagnetic and ferromagnetic states, and the contribution from the paramagnetic state is dominant in the initial state.
Next, we discuss the properties of the phase transition point in more detail.
To this end, we perform another finite-size scaling analysis for all $\gamma$ shown in Fig.~\ref{fig5}. For the result of $\gamma/w=8.0$, see Fig.~\ref{fig6}. 
We employ the critical exponent $\eta$ of the equilibrium classical XY model ($\eta=0.038$~\cite{PhysRevB.63.214503}), which represnts the anomalous dimension of the spin operator (order parameter).
The rescaled spin structure factor shows scale invariance at the phase transition time, which means that the phase transision point is actually a critical point.
We note that $\eta$ is so small that we can check the consistency, but cannot determine $\eta$ solely from the Monte Carlo data.

\section{Discussion} How can we understand the universality class of this dynamical phase transition? 
The system has $\mathrm{SU}(2)$ symmetry because the Hamiltonian~\eqref{eq:nonhermite} commutes with the spin operator $\bm{S}=\sum_{\bm r}\bm S_{\bm r}$. However, the initial N\'{e}el state explicitly breaks the $\mathrm{SU}(2)$ spin symmetry to SO($2$)$\simeq$U($1$), and remaining symmetry is further spontaneously broken during the time evolution.
From the viewpoint of the projector quantum Monte Carlo, the explicit symmetry breaking by a trial wave function is negligible in the large projection-time limit, but is relevant at finite time.
Therefore the symmetry breaking pattern of the dynamical phase transition is expected to be
\be
{\rm U(1)} \to \varnothing .
\ee
To relate this with the {\it equilibrium} classical XY universality, let us consider a unitary transformation $c_{\bm r\sigma}\rightarrow i^{x+y+z} c_{\bm r\sigma}$, and obtain the non-Hermitian Hamiltonian $H_{\rm new} \equiv i H_{\rm eff}$ as
\bea
H_{\rm new} = -w\sum_{\bm r,j,\sigma}\left[  c^\dagger_{\bm r\sigma}c_{\bm r+\hat{j}\,\sigma}-c^\dagger_{\bm r+\hat{j}\,\sigma}c_{\bm r\sigma}\right]
+\gamma\sum_x  c^\dagger_{\bm r\up}c_{\bm r\up} c^\dagger_{\bm r\down}c_{\bm r\down}  .\;\;\;\;
\eea
The real-time evolution $e^{-iH_{\rm eff} t}$ is mapped to the imaginary-time one $e^{-H_{\rm new} t}$. 
The strong coupling expansion of $H_{\rm new}$ follows the standard perturbation theory, and results in the ferromagnetic Heisenberg model $H_{\rm spin}$ at second order (see Appendix~\ref{sec:strongcoupling})~\cite{PhysRevLett.124.147203}. 
Therefore, Eq.~\eqref{eq:expectation} is reduced to $\langle\Psi(0)|e^{-t H_{\rm spin}}\hat{O}e^{-t H_{\rm spin}}|\Psi(0)\rangle/\langle\Psi(0)|e^{-2t H_{\rm spin}}|\Psi(0)\rangle$. 
From this expression, we expect that our dynamic universality class may be related to the equilibrium one although the expectation value is taken by a pure state, not the trace average.
In equilibrium spontaneous symmetry breaking, the finite-size scaling in the ordered state is linear for the U($1$) symmetry, while it is quadratic for the SU($2$) symmetry~\cite{Staudt_2000}.
Therefore, with the help of the strong coupling expansion analysis, our finite-size scaling analysis shown in Fig.~\ref{fig4} implies that the dynamical phase transition belongs to the XY universality class.
The finite-size scaling analysis shown in Fig.~\ref{fig6} indeed supports this understanding.
Interestingly, the universality class of our novel dynamical phase transition depends on the symmetry of the initial state as well as that of the Hamiltonian,
which is in stark contrast to the universality class of equilibrium phase transitions.
For example, we chose the N\'{e}el state as an initial state, and then our results should be compared with the Heisenberg model with the fixed $S^z$, which has the same universality class as the classical XY model.
Instead, if we choose the SU($2$) symmetric state as an initial state, the universality class may be changed to the Heisenberg universality, which can be confirmed by the quantum Monte Carlo simulation.

\section*{Acknowledgment}

This work was supported by JSPS KAKENHI Grant Numbers 19K03841, 21H01007, and 21H01084.
The numerical calculations were carried out on Yukawa-21 at YITP in Kyoto University, and on cluster computers at iTHEMS in RIKEN.

\appendix

\section{Quantum Monte Carlo}
\label{sec:QMC}
We here detail an application of the fermion quantum Monte Carlo to the real-time problem.
Using the total number of the fermions $N=\sum_{\bm{r},\sigma}c^\dag_{\bm{r}\sigma} c_{\bm{r}\sigma}$, we rescale the wave function as
$|\Psi(t)\rangle=e^{-\frac{\gamma}{2}Nt}|\tilde{\Psi}(t)\rangle$. 
Since $N$ is conserved during time evolution, that is, it is a classical number fixed by the initial condition, an expectation value of a physical operator given in Eq.~($6$) in the main text does not change under the transformation.
The time evolution of $|\tilde{\Psi}(t)\rangle$ obeys
\bea
i\frac{d}{dt}|\tilde{\Psi}(t)\rangle = \left(H_{\rm eff}+i\frac{\gamma}{2}\sum_{\bm r,\sigma}c^\dagger_{\bm r\sigma}c_{\bm r\sigma}\right)|\tilde{\Psi}(t)\rangle
\equiv \tilde{H}_{\rm eff}|\tilde{\Psi}(t)\rangle ,
\eea
where $H_{\rm eff}$ is given by Eq.~\eqref{eq:nonhermite} in the main text.
The transformation is essential for the sign-free auxiliary-field quantum Monte Carlo.
Below we always use the latter representation, and omit the tilde index for notational simplicity.

By using the quantum Monte Carlo, we numerically solve the Schr{\"o}dinger equation 
\bea
i\frac{d}{dt}|\Psi(t)\rangle = H_{\rm eff}|\Psi(t)\rangle ,
\eea
with the non-Hermitian Hamiltonian
\bea
H_{\rm eff} =
 -w\sum_{\bm r,j,\sigma}\left[ c^\dagger_{\bm r\sigma}c_{\bm r+\hat{j}\,\sigma}+c^\dagger_{\bm r+\hat{j}\,\sigma}c_{\bm r\sigma}\right]
-i\gamma\sum_{\bm r}  c^\dagger_{\bm r\up}c^\dagger_{\bm r\down}c_{\bm r\down}c_{\bm r\up}  
+i\frac{\gamma}{2}\sum_{\bm r,\sigma}c^\dagger_{\bm r\sigma}c_{\bm r\sigma} .
\label{eq:nonhermite_app}
\eea
Using the second-order Suzuki-Trotter decomposition, we write the persistent probability as
\bea
\langle\Psi(t)|\Psi(t)\rangle = \langle\Psi(0)|e^{iH^\dagger_{\rm eff}t}e^{-iH_{\rm eff}t}|\Psi(0)\rangle
=\langle\Psi(0)|\prod_m \cU_m^\dagger \prod_n \cU_n|\Psi(0)\rangle ,
\eea
where
\bea
\cU_n &=&e^{-i\Delta t K/2}e^{-i\Delta t U}e^{-i\Delta t K/2},
\\
K &=&-w\sum_{\bm r,j,\sigma}\left[ c^\dagger_{\bm r\sigma}c_{\bm r+\hat{j}\,\sigma}+c^\dagger_{\bm r+\hat{j}\,\sigma}c_{\bm r\sigma}\right],
\\
U &=&i\gamma\sum_{\bm r}\left[ - c^\dagger_{\bm r\up}c_{\bm r\up}c^\dagger_{\bm r\down}c_{\bm r\down}+\frac{1}{2}(c^\dagger_{\bm r\up}c_{\bm r\up}+c^\dagger_{\bm r\down}c_{\bm r\down})\right],\;\;\;\;
\eea
and $\Delta t=t/N_t$ with $N_t$ being the number of the Suzuki-Trotter step.
Each component of $U$ can be rewritten by introducing the auxiliary binary field $s$ as
\bea
e^{-\gamma\Delta t \left(c^\dagger_{\bm r\up}c_{\bm r\up}-\frac{1}{2}\right)\left(c^\dagger_{\bm r\down}c_{\bm r\down}-\frac{1}{2}\right)+\frac{\gamma\Delta t}{4}}
= \frac{e^{\frac{\gamma\Delta t}{2}}}{2}\sum_{s=\pm 1}e^{igs \left(c^\dagger_{\bm r\up}c_{\bm r\up}+c^\dagger_{\bm r\down}c_{\bm r\down}-1\right)} ,
\eea
where $\cos(g)=e^{-\gamma\Delta t/2}$.
Therefore, the nonunitary evolution $e^{-iH_{\rm eff}t}$ is reformulated as the unitary evolution under the space-time binary disorder.  
Now we have~\cite{Assaad2008}
\bea
\langle\Psi(t)|\Psi(t)\rangle = {\cal N}\sum_{\{s(n,\bm r)\}}e^{-ig\sum_{n,\bm r}s(n,\bm r)}\prod_\sigma{\rm det}\left[P_\sigma^\dagger B_{2N_t}\cdots B_1P_\sigma\right] ,\;\;\;\;
\eea
where ${\cal N}=(e^{\frac{\gamma\Delta t}{2}}/2)^{2N_tV}$ is the normalization factor with $V$ being the total number of the lattice sites.
The $V\times V$ matrix $B_n$ is given as
\bea
 B_n = 
\begin{cases}
 e^{-i\Delta t k/2}e^{u_n}e^{-i\Delta t k/2} & (n \le N_t)\\
 e^{i\Delta t k/2}e^{u_n}e^{i\Delta t k/2} & (n > N_t),
\end{cases}
\eea
where $k$ and $u_n$ are the matrix representation of $K\equiv \sum_{\bm r,\bm r^\prime,\sigma} c^\dagger_{\bm r\sigma}k_{\bm r\bm r^\prime}c_{\bm r^\prime\sigma}$ and $\sum_{\bm r}igs({n,\bm r}) (c^\dagger_{\bm r\up}c_{\bm r\up}+c^\dagger_{\bm r\down}c_{\bm r\down}) \equiv \sum_{\bm r,\bm r^\prime,\sigma} c^\dagger_{\bm r\sigma} [u_n]_{\bm r\bm r^\prime}c_{\bm r^\prime\sigma}$.
We here used two facts: (I) The initial state is a direct product state of each spin state $|\Psi(0)\rangle=|\Psi_\up(0)\rangle\otimes |\Psi_\down(0)\rangle$. (II) Each spin state is expressed by a Slater determinant with the $V\times V/2$ rectangular matrix $P_\sigma$:
\bea
|\Psi_\sigma(0)\rangle
=\prod_{\bm r^\prime}^{V/2}c^\dagger_{\bm r^\prime\sigma} |0\rangle
=\prod_{\bm r^\prime}^{V/2}\sum^V_{\bm r}c^\dagger_{\bm r\sigma}[P_\sigma]_{\bm r\bm r^\prime} |0\rangle ,
\eea
where $|0\rangle$ is the Fock vacuum, and $\bm r^\prime$ runs for even (odd) sites for the $\up$- ($\down$-) component.

We consider the particle-hole transformation only for the $\down$-component: $c_{\bm r\down}\rightarrow c^\dagger_{\bm r \down}$. Then, occupied and unoccupied states are swapped for the $\down$-component, so that $|\Psi_\down(0)\rangle=|\Psi_\up(0)\rangle$, and $P_\down=P_\up$ (Note that we choose the N{\'e}el state as an initial state).
Also, the transfer matrix is transformed as $e^{-ig\sum_{\bm r} s(n,\bm r)}B_n\rightarrow B_n^*$, and thus the integrand of $\langle\Psi(t)|\Psi(t)\rangle$ is actually positive definite:
\bea
\langle\Psi(t)|\Psi(t)\rangle 
={\cal N}\sum_{s(n,\bm r)}\left|{\rm det}\left[P_\up^\dagger B_{2N_t}\cdots B_1P_\up\right]\right|^2 .
\eea
The summation of the auxiliary fields can be evaluated on the basis of the importance sampling, and the physical observable can by ensemble average (with the help of the Wick theorem) as is commonly done in the projector quantum Monte Carlo (See, e.g., Ref.~\cite{Assaad2008}).

\section{Strong coupling expansion}
\label{sec:strongcoupling}
The real-time evolution of the original model $H_{\rm eff}$ can be mapped to the imaginary-time one of the non-Hermitian Hubbard model with the sign-asymmetric hopping
\bea
H_{\rm new} = -w\sum_{\bm r,j,\sigma}\left[  c^\dagger_{\bm r\sigma}c_{\bm r+\hat{j}\,\sigma}-c^\dagger_{\bm r+\hat{j}\,\sigma}c_{\bm r\sigma}\right]
+\gamma\sum_x  c^\dagger_{\bm r\up}c_{\bm r\up} c^\dagger_{\bm r\down}c_{\bm r\down}  .\;\;\;\;
\eea
We here describe the strong coupling expansion of $H_{\rm new}$.

In the infinite coupling limit $\gamma \to \infty$, hopping terms are negligible, so an effective Hilbert space is the same as that of the Hermitian-Hubbard model, that is, the half-filling Hilbert space with no double occupancy.
Since hopping terms change the occupancy, there is no first-order correction $O(w/\gamma)$.
At the second order, the non-Hermitian Hamiltonian in Eq.~(10) in the main text is reduced to 
\be
H_{\rm new}= E_0+P K'\frac{1}{E_0-H_0}K'P ,
\label{eq:second}
\ee
where $E_0=0$, $P$ is a projector to the states with no double occupancy, and 
\bea
K' &=&-w\sum_{\bm r,j,\sigma}\left[ c^\dagger_{\bm r\sigma}c_{\bm r+\hat{j}\,\sigma}-c^\dagger_{\bm r+\hat{j}\,\sigma}c_{\bm r\sigma}\right],
\\
H_0 &=&\gamma\sum_{\bm r} c^\dagger_{\bm r\up}c_{\bm r\up}c^\dagger_{\bm r\down}c_{\bm r\down} .
\eea
By taking care of the sign of hopping terms, we compute Eq.~\eqref{eq:second}, and obtain 
\bea
H_{\rm new}
&=& -\frac{w^2}{\gamma} \sum_{\bm r,j} \Bigl[
 c^\dagger_{\bm r\up}c_{\bm r+\hat{j}\,\up}c^\dagger_{\bm r+\hat{j}\,\down}c_{\bm r\down}
+c^\dagger_{\bm r+\hat{j}\,\down}c_{\bm r\down}c^\dagger_{\bm r\up}c_{\bm r+\hat{j}\,\up}
\notag \\
&&+c^\dagger_{\bm r\down}c_{\bm r+\hat{j}\,\down}c^\dagger_{\bm r+\hat{j}\,\up}c_{\bm r\up}
+c^\dagger_{\bm r+\hat{j}\,\up}c_{\bm r\up}c^\dagger_{\bm r\down}c_{\bm r+\hat{j}\,\down}
\notag \\
&&+c^\dagger_{\bm r\up}c_{\bm r+\hat{j}\,\up}c^\dagger_{\bm r+\hat{j}\,\up}c_{\bm r\up}
+c^\dagger_{\bm r+\hat{j}\,\up}c_{\bm r\up}c^\dagger_{\bm r\up}c_{\bm r+\hat{j}\,\up}
\notag \\
&&+c^\dagger_{\bm r\down}c_{\bm r+\hat{j}\,\down}c^\dagger_{\bm r+\hat{j}\,\down}c_{\bm r\down}
+c^\dagger_{\bm r+\hat{j}\,\down}c_{\bm r\down}c^\dagger_{\bm r\down}c_{\bm r+\hat{j}\,\down}
\Bigr]
\notag \\
&=& - \frac{4w^2}{\gamma} \sum_{\bm r,j} \left(\bm S_{\bm r}\cdot\bm S_{\bm r+\hat{j}}- \frac{1}{4}\right) .
\eea
We used $c^\dagger_{\bm r\up}c_{\bm r\up}+c^\dagger_{\bm r\down}c_{\bm r\down}=1$, which holds for the half-filling Hilbert space with no double occupancy, and
\bea
&&2\left(S^x_{\bm r}S^x_{\bm r+\hat{j}}+S^y_{\bm r}S^y_{\bm r+\hat{j}}\right) =
c^\dagger_{\bm r\up}c_{\bm r\down}c^\dagger_{\bm r+\hat{j}\,\down}c_{\bm r+\hat{j}\,\up}
+c^\dagger_{\bm r\down}c_{\bm r\up}c^\dagger_{\bm r+\hat{j}\,\up}c_{\bm r+\hat{j}\,\down} ,
 \\
&&4S^z_{\bm r}S^z_{\bm r+\hat{j}} =
2\left(c^\dagger_{\bm r\up}c_{\bm r\up}c^\dagger_{\bm r+\hat{j}\,\up}c_{\bm r+\hat{j}\,\up}+c^\dagger_{\bm r\down}c_{\bm r\down}c^\dagger_{\bm r+\hat{j}\,\down}c_{\bm r+\hat{j}\,\down}\right)
-c^\dagger_{\bm r\up}c_{\bm r\up}
-c^\dagger_{\bm r\down}c_{\bm r\down} .
\eea
Therefore, within the second order $O(w^2/\gamma^2)$ of the strong coupling expansion, the real-time problem of the original model $H_{\rm eff}$ is reduced to the imaginary-time one of the ferromagnetic Heisenberg model. 

\bibliographystyle{ptephy}
\bibliography{nonhermite}

\end{document}